\documentclass[prb,twocolumn,showpacs]{revtex4}
\usepackage{graphicx}
\usepackage{bm}
\usepackage{amssymb,amsmath}
\usepackage[usenames]{color}
\usepackage{epsfig}
\usepackage{hyperref}
\usepackage{subfigure}
\usepackage[sans]{dsfont}
\hypersetup{colorlinks=true, citecolor=blue,
linkcolor=blue,urlcolor=blue }

\begin{document}

\title{ Dynamic renormalization group analysis of propagation of
elastic waves in two-dimensional heterogeneous media}

\author{Reza Sepehrinia\textsuperscript{1}, Alireza Bahraminasab\textsuperscript{1,2}, Muhammad
Sahimi\textsuperscript{3,*} and M. Reza Rahimi Tabar\textsuperscript{1,4,5} \\
\textsuperscript{1}\textit{Department of Physics, Sharif University of
Technology, Tehran 11365-9161, Iran} \\
\textsuperscript{2}\textit{Department of Physics, Lancaster University,
Lancaster LA1 4YB, United Kingdom} \\
\textsuperscript{3}\textit{Mork Family Department of Chemical
Engineering Materials Science, University of Southern California,
\\ Los Angeles,
California 90089-1211, USA} \\
\textsuperscript{4}\textit{Observatoire de la C\^{o}te d'Azur, CNRS UMR 6529, Bo\^{\i}te Postale 4229, 06304 Nice Cedex 4, France} \\
\textsuperscript{5}\textit{Carl Institute of Physics, von Ossietzky
University, D-26111 Oldenburg, Germany }}

\begin{abstract}
We study localization of elastic waves in two-dimensional
heterogeneous solids with randomly distributed Lam\'e coefficients,
as well as those with long-range correlations with a power-law
correlation function. The Matin-Siggia-Rose method is used, and the
one-loop renormalization group (RG) equations for the the coupling
constants are derived in the limit of long wavelengths. The various
phases of the coupling constants space, which depend on the value
$\rho$, the exponent that characterizes the power-law correlation
function, are determined and described. Qualitatively different
behaviors emerge for $\rho<1$ and $\rho>1$. The Gaussian fixed point
(FP) is stable (unstable) for $\rho<1$ ($\rho>1$). For $\rho<1$
there is a region of the coupling constants space in which the RG
flows are toward the Gaussian FP, implying that the disorder is
irrelevant and the waves are delocalized. In the rest of the
disorder space the elastic waves are localized. We compare the
results with those obtained previously for acoustic wave propagation
in the same type of heterogeneous media, and describe the
similarities and differences between the two phenomena.
\end{abstract}

 \maketitle

\begin{center}
{\bf I. INTRODUCTION}
\end{center}

Ever since the discovery of electron localization,$^1$ much
attention has been devoted to this phenomenon, since it is not only
of fundamental scientific interest, but also has much practical
importance. There is now extensive experimental evidence for the
localization phenomenon in disordered materials.$^{2,3}$ On the
theoretical side, the problem has been studied for decades by
several analytical methods, ranging from the scaling theory$^4$ to
the self-consistent perturbation theory.$^{5,6}$ In addition,
numerical simulations using such techniques as the transfer-matrix
method and the statistics of energy levels have been used to verify
the predictions of the analytical results.

The development of the one-parameter scaling theory$^4$ of electron
localization in terms of the concepts of critical phenomena suggests
that, the problem can be reformulated by using an effective field
theory which, when done, leads to the so-called $\sigma$ model which
is a nonlinear model. Wegner$^7$ proposed such a description of
disordered conductors. Also noteworthy among the theoretical
developments is the work of Efetov {\it et al.},$^8$ who proposed
the supersymmetric approach, now used widely. The renormalization
group (RG) approach, one of the most powerful methods in statistical
physics, has also been used to examine the critical properties of
the resulting  effective field theory.$^9$ The RG approach leads to
a set of equations for the coupling constants, such as the
diffusivity and conductance of the disordered materials under study.
The main prediction of all of these approaches is that, for space
dimensions $d>2$, there is a transition from the localized to
extended states, so that the lower critical dimension of the
localization phenomenon is, $d_c=2$. However, despite convincing
numerical evidence for the validity of this prediction,$^{10}$ the
exponent $\nu$ that characterizes the power-law behavior of the
localization length $\xi$ near the phase transition, $\xi\propto
|W-W_c|^{-\nu}$ (where $W_c$ is the critical value of the disorder
intensity $W$), predicted by the RG method, is not in agreement with
the numerical results. A possible explanation for this discrepancy
is that, some of the terms that are neglected in the construction of
the field theory may actually be relevant to the RG analysis.

Another approach to the field-theoretic description of the problem
is based on the method first developed by Martin, Siggia, and Rose
(MSR),$^{11}$ by which one constructs an {\it effective action} (see
below) based on the governing stochastic equation of motion for the
phenomenon under study. The MSR approach is well developed for
critical phenomena far from equilibrium,$^{11}$ and has been
extensively used to study various dynamical critical phenomena, such
as those that are described by the Langevin equation, or the driven
interface phenomena, such as surface growth and stochastic
hydrodynamics. The main advantages of the MSR method are that, it
provides an exact generating functional, and that one needs no
approximation in order to obtain the effective action. This is the
method that we use in the present paper.

An important implication of the wave characteristics of electrons is
that, the localization phenomenon may also occur in propagation of
the {\it classical} waves in disordered media. However, unlike the
problem of electron localization in strongly disordered materials,
classical waves, such as seismic waves,$^{12,13}$ do not interact
with each other and, therefore, their propagation in heterogeneous
media provides an ideal model for studying the phenomenon of
localization of the classical waves. Moreover, along with the work
on electronic transport in disordered materials, parallel work has
been carried out on localization properties of classical waves in
disordered media that describe the phonons that are responsible for
heat transport in solids.$^{14}$

Although waves that are described by scalar equations have been used
for describing phonons in disordered materials, a more suitable
continuum description of the phenomenon is through propagation of
elastic waves. Due to the presence of different polarizations and
the coupling between them (mode conversion), propagation of elastic
waves in disordered solids constitutes a complex set of
phenomena.$^{15,16}$ Because of this complexity, there have been
relatively few studies in the literature dealing with propagation of
elastic waves in disordered solid.$^{17}$ In particular,
localization of special types of elastic waves has been studied in
the past, ranging from surface elastic (Rayleigh) waves,$^{18}$ to
transverse deflections of a beam,$^{19}$ and coherent backscattering
and multiple scattering.$^{20,21}$

At the same time, understanding how elastic waves propagate at very
large scales, particularly in highly heterogeneous media such as
rock, is fundamental to a host of other important problems, such as
earthquakes, underground nuclear explosions, the morphology and
content of oil and gas reservoirs, oceanography, and materials
sciences.$^{12}$ For example, seismic wave propagation and
reflection are used to not only estimate the hydrocarbon content of
a potential oil or gas field and gain insight into its morphology,
but also to image structures located over a wide area, ranging from
the Earth's near surface to the deeper crust and upper
mantle.$^{22}$

The purpose of the present paper is to study the effect of
heterogeneities, represented by spatial distributions of the local
elastic constants, on elastic wave propagation in disordered media,
such as rock, which represents a highly heterogeneous natural
material. Recently, extensive experimental data for the spatial
distributions of the local elastic moduli, the density, and the wave
velocities in several large-scale porous rock formations, both off-
and onshore, were analyzed.$^{23}$ The analysis provided strong
evidence for the existence of long-range correlations in the spatial
distributions of the measured quantities, characterized by a
power-law correlation function. The existence of such correlations
in the data provided the impetus for the present study and motivated
an important question that we address in the present paper: how do
large-scale heterogeneities and long-range correlations affect
elastic wave propagation in disordered media? Another question that
we address in the present paper is whether, in the presence of the
heterogeneities, the elastic waves can be delocalized. By
localization we mean a situation in which, over finite length scales
(which can, however, be large), the waves' amplitude decays and
essentially vanishes.

Localization of elastic waves in rock would imply, for example, that
seismic exploration yields useful information only over distances
$r$ from the explosion's site that are of the order of the
localization length $\xi$. Thus, if, for example, $\xi$ is on the
order of a few kilometers, but the linear size of the area for which
a seismic exploration is done is significantly larger than $\xi$,
then, seismic recordings can, at best, provide only partial
information about the area. Localization of elastic waves also
implies that, if the stations that collect data for seismic waves
that are emanated from an earthquake in rock are farther from the
earthquake's hypocenter than $\xi$, no useful information on the
seismic activity prior to and during the earthquake can be gleaned
from the data$^{24}$.

We use a field-theoretic formulation to study propagation of elastic
waves in two-dimensional (2D) disordered media in which the Lam\'e
coefficients are spatially distributed. Our approach is based on the
MSR method.$^{11}$ We calculate the one-loop $\beta$ functions (see
below) for both spatially random and power-law correlated
distribution$^{23}$ of the local elastic constants. Although our
work is primarily motivated by the analysis of experimental data for
the spatial distribution of elastic constants of rock at large
scales,$^{23}$ the results presented in this paper are general and
applicable to any solid material in which the local elastic
constants follow the statistics of the distributions that we
consider. The present paper represents the continuation of our
previous work$^{25,26}$ in which we studied acoustic wave
propagation in the same type of heterogeneous media. We will compare
the results with those obtained previously for propagation of
acoustic waves.

The rest of this paper is organized as follows. In Sec. II the model
is described and the governing equations are presented. Section III
describes the field-theoretic description of the elastic wave
equation, and the development of the MSR formulation for the
propagation of the waves in heterogeneous media. In Sec. IV the
perturbative RG calculations, based on the MSR action, are carried
out and the results are analyzed. In  Sec. V we compare the results
with those obtained previously$^{25,26}$ for propagation of acoustic
waves in the same type of heterogeneous media that we consider in
the present paper. The paper is summarized in Sec. VI.

\begin{center}
{\bf II. THE MODEL AND GOVERNING EQUATIONS}
\end{center}

To analyze propagation of elastic waves in a disordered medium, we
begin with the equation of motion of an elastic medium with the mean
density $m$,
\begin{eqnarray}\label{wave equation}
m\frac{\partial^2 u_i}{\partial t^2}=\partial_j\sigma_{ij}\;,
\end{eqnarray}
where $u_i$ is the displacement in the $i$th direction, and
$\sigma_{ij}$ the $ij$th component of the stress tensor
$\mbox{\boldmath$\sigma$}$. As usual, $\sigma_{ij}$ is expressed in
terms of the strain tensor,
\begin{equation}
\sigma_{ij}({\bf x})=2\mu({\bf x})u_{ij}+\lambda({\bf
x})u_{kk}\delta_{ij}\;.
\end{equation}
For small deformations, the strain tensor is given by,
\begin{equation}
u_{ij}=\frac{1}{2}(\partial_i u_j+\partial_ju_i)\;,
\end{equation}
where $\lambda$ and $\mu$ are the Lam\'e coefficients. For
simplicity, we take the two Lam\'e coefficients to be equal, but the
main results of the paper presented below will not change if they
are unequal, but follow the same type of statistical distributions.
Hence, we write,
\begin{equation}
\mu({\bf x})=\lambda({\bf x})=\lambda_0+\eta({\bf x})\;,
\end{equation}
where $\lambda_0=\langle\lambda({\bf x})\rangle$, with
$\langle\cdot\rangle$ representing a spatial averaging. We assume
that $\eta({\bf x})$, the fluctuating part of the Lam\'e
coefficients, is a Gaussian random process. Thus, in performing the
spatial average over the disorder we use a probability distribution
of the form
\begin{eqnarray}\label{Gaussian}
P[\eta({\bf x})]\propto \exp\left[-\int d{\bf x}d{\bf x'}\eta({\bf
x}) D({\bf x}-{\bf x'})\eta({\bf x'})\right]\;,
\end{eqnarray}
where $D({\bf x})$ is the inverse of the correlation function
$C({\bf x})$. The disorder that we include in the model consists of
two parts. One is (random) $\delta-$correlated, while the second
part is characterized by a power-law correlation function. Hence,
the overall correlation function of the spatial distribution of the
disorder is given by
\begin{widetext}\begin{eqnarray}
\langle\eta({\bf x})\eta({\bf x}')\rangle=2C({\bf x}-{\bf
x}')=2D_0\delta^d ({\bf x}-{\bf x}')+2D_\rho|{\bf x}-{\bf
x}'|^{2\rho-d}\;,
\end{eqnarray}
\end{widetext} in which $D_0$ and $D_\rho$ are, respectively, the
strengths of the disorder for the random and the power-law
correlated parts, $C({\bf x}-{\bf x'})$ satisfies the following
condition
\begin{equation}
\int d{\bf x''}C({\bf x}-{\bf x''})D({\bf x''}-{\bf x'})=\delta({\bf
x}- {\bf x'})\;,
\end{equation}
and $d$ is the spatial dimension ($d=2$ in this paper). Note that,
in 2D, $\rho=H+1$, with $H$ being the Hurst exponent.

A Gaussian distribution of the form (\ref{Gaussian}) gives rise to
quadratic couplings in the interaction part of the action defined
below. Moreover, the Gaussian distribution (5) may include a tail of
inadmissible negative values of the Lam\'e coefficients. In
principle, the unphysical tail can be removed by introducing a
modified probability distribution function which, however, would
produce couplings of higher order in the action. But, interactions
of orders higher than quadratic are irrelevant in the RG analysis
and, therefore, can be ignored.

We now take the Fourier transform of Eq. (\ref{wave equation}) with
respect to the time variable, which yields the governing equation
for a monochromatic wave with angular frequency $\omega$,
\begin{eqnarray}\label{frequency}
\partial_j\sigma_{ij}+\omega^2mu_i\equiv \lambda_0{\cal L}_{ij}u_j=0\;.
\end{eqnarray}
Here, ${\cal L}$ is a $2\times 2$ differential matrix operator (see
below).

\begin{center}
{\bf III. FIELD-THEORETIC REPRESENTATION OF THE ELASTIC WAVE
EQUATION}
\end{center}

Using the formalism developed by De Dominicis and Peliti$^{27}$ (see
also Hochberg {\it et al.}$^{28}$), one obtains a MSR generating
functional that corresponds to the (Fourier-transformed) wave
equation (\ref{frequency})
\begin{widetext}
\begin{eqnarray}\label{generating function}
P[u_i^R,u_i^I] &=& \frac{1}{\cal N}\int [{\cal D}\eta] [{\cal
D}\{u_i^R,u_i^I\}]\delta\left({\cal L}_{1j}u_j^R\right)\delta
\left({\cal L}_{2j}u_j^R\right)\delta\left({\cal
L}_{1j}u_j^I\right)\delta
\left({\cal L}_{2j}u_j^I\right)\nonumber\\
&\times & J\left(\frac{\partial{\cal L}{\bf u}^R}{\partial{\bf
u}^R}\right)J \left(\frac{\partial{\cal L}{\bf u}^I}{\partial{\bf
u}^I}\right) \exp\left[-\int d{\bf x}d{\bf x'}\eta({\bf x})D({\bf
x}-{\bf x'})\eta({\bf x'}) \right]\;.
\end{eqnarray}
\end{widetext} Here, superscripts $R$ and $I$ indicate, respectively,
the real and imaginary parts of the solution of the wave equation,
$J$ is the Jacobian and, ${\cal N}$ is a normalization constant. The
Jacobian for the transformation, ${\bf u}\to {\cal L}{\bf u}$, is
expressed as a Grassman integral over the anticommuting fields
$\chi_i,\chi_i^*$,
\begin{widetext}
\begin{eqnarray}\label{Jacobian}
J=\int {\cal D}\{\chi^*_i,\chi_i\}\exp\left\{\int dx\left[
\begin{array}{cc}
\chi^*_1({\bf x}) & \chi^*_2({\bf x})
\end{array}\right]\left(
\begin{array}{cc}
{\cal L}_{11} & {\cal L}_{12}\\
{\cal L}_{21} & {\cal L}_{22}
\end{array}\right)\left[
\begin{array}{c}
\chi_1({\bf x}) \\
\chi_2({\bf x})
\end{array}\right]\right\}\;.
\end{eqnarray}
\end{widetext} We now introduce two other auxiliary fields,
$\tilde{u}_i^R$ and $\tilde{u}_i^I$, in order to express the
$\delta$-functions in Eq. (\ref{generating function}) as Fourier
transforms. Then, substituting Eq. (\ref{Jacobian}) in Eq.
(\ref{generating function}) and integrating out $\eta$, (by
performing a Gaussian integration) leads to an effective MSR action
$S_e$ with the following form
\begin{widetext}
\begin{eqnarray}
S_e &=& S_0+S_I\\
S_0 &=& \int d{\bf x}\sum_{a=R,I}\left[i\;\tilde{\bf u}^a({\bf x})
\cdot{\cal L}_0{\bf u}^a({\bf x})+\mbox{\boldmath$\chi$}^{a*}({\bf
x})\cdot
{\cal L}_0\mbox{\boldmath$\chi$}^a({\bf x})\right]\\
S_I &=& \int d{\bf x}d{\bf
x}'\left[\sum_{a=R,I}i\;\partial_j\tilde{u}^a_i (2\partial_i
u^a_j+2\partial_j u^a_i+\partial_k u^a_k\delta_{ij})+\partial_j
\chi^{a*}_i(2\partial_i\chi^a_j+2\partial_j\chi^a_i+\partial_k
\chi^a_k\delta_{ij})\right]_{\bf x}\\
&\times& \frac{C({\bf x}-{\bf
x}')}{2\lambda_0^2}\left[\sum_{a=R,I}i\;
\partial_j\tilde{u}^a_i(2\partial_i u^a_j+2\partial_j u^a_i+\partial_k
u^a_k\delta_{ij})+\partial_j\chi^{a*}_i(2\partial_i\chi^a_j+2\partial_j
\chi^a_i+\partial_k\chi^a_k\delta_{ij})\right]_{{\bf x}'}\;,
\end{eqnarray}
\end{widetext} where the subscripts {\bf x} and ${\bf x}'$ indicate
where the quantities are evaluated at. The explicit form of the
matrix ${\cal L}_0$ is given by
\begin{eqnarray*}
{\cal L}_0=\left(\begin{array}{cc}
3\partial_x^2+\partial_y^2+\omega^2/\lambda_0 & 2\partial_x\partial_y\\
2\partial_x\partial_y &
\partial_x^2+3\partial_y^2+\omega^2/\lambda_0
\end{array}\right)\;.
\end{eqnarray*}

We now write down the action in the Fourier space and, then,
introduce a change of the basis to decouple the free propagator into
two components, the longitudinal and transverse propagators. To do
so, we use a transformation ${\bf A}\to {\bf U A}$ in order to
diagonalize the matrix ${\cal L}_0$ in the Fourier space, where it
has the following form,
\begin{eqnarray*}
{\cal L}_0=\left(\begin{array}{cc}
-3k_x^2-k_y^2+\omega^2/\lambda_0 & -2k_xk_y\\ \\
-2k_xk_y & -k_x^2-3k_y^2+\omega^2/\lambda_0
\end{array}\right)
\end{eqnarray*}
with following eigenvectors
\begin{displaymath}
|1\rangle=\frac{1}{k}\left(\begin{array}{c}
k_x \\
k_y
\end{array}\right)\;,
\hspace{1cm} |2\rangle=\frac{1}{k}\left(\begin{array}{c}
-k_y \\
k_x
\end{array}\right)\;.
\end{displaymath}
The corresponding eigenvalues are ($\omega^2/\lambda_0-k^2$) and
($\omega^2/\lambda_0-3k^2$), respectively. The two eigenvalues
represent the dispersion relations of the transverse and
longitudinal waves which propagate in a uniform medium with the
phase velocities, $v_t=\sqrt{\lambda_0}$ and
$v_l=\sqrt{3\lambda_0}$. Using the eigenvectors, the transformation
matrix {\bf U} is given by
\begin{eqnarray}\label{U}
{\bf U}=\frac{1}{k}\left(\begin{array}{cc}
k_x & -k_y\\
k_y & k_x
\end{array}\right)\;.
\end{eqnarray}
By applying the transformation {\bf U}, we finally obtain
\begin{widetext}
\begin{eqnarray}\label{action}
S_0&=&\int_k  \sum_{a=R,I}\left[i\tilde{\bf u}^a(-k)\cdot{\cal
L}_0^d {\bf u}^a(k)+\mbox{\boldmath$\chi$}^{a*}(-k)\cdot{\cal L}_0^d
\mbox{\boldmath$\chi$}^a(k)\right],\nonumber\\
S_I &=&\int_{k,p_i}\left[\sum_{a=R,I} i\tilde{\bf
u}^a(p_1)\cdot{\cal L}_I {\bf
u}^a(p_2)+\mbox{\boldmath$\chi$}^{a*}(p_1)\cdot{\cal
L}_I\mbox{\boldmath$ \chi$}^a(p_2)\right]  \left[g_0\delta(\sum_i
p_i)+g_\rho k^{-2\rho}\delta(p_1+p_2-k)
\delta(p_3+p_4+k)\right]\nonumber\\
 && \ \ \ \ \ \ \times  \left[\sum_{a=R,I} i\tilde{\bf
u}^a(p_3)\cdot{\cal L}_I{\bf u}^a
(p_4)+\mbox{\boldmath$\chi$}^{a*}(p_3)\cdot{\cal
L}_I\mbox{\boldmath$\chi$}^a (p_4)\right]\;,
\end{eqnarray}
 with
\begin{displaymath}
{\mathcal L}_0^d=\left(\begin{array}{cc}
-k^2+\omega^2/\lambda_0 & 0\\
0 & -3k^2+\omega^2/\lambda_0
\end{array}\right)\;,
\hspace{5mm}{\cal L}_I=\left(\begin{array}{cc}
A(p_1,p_2)  & -C(p_1,p_2)\\
C(p_1,p_2)  & B(p_1,p_2)
\end{array}\right)\;,
\end{displaymath}
\end{widetext} where,
\begin{eqnarray*}
& & A=c\left[3(p_1\cdot p_2)^2+|p_1\times p_2|^2\right]\;,\\
& & B=c\left[(p_1\cdot p_2)^2-|p_1\times p_2|^2\right]\;,\\
& & C=c\left[2(p_1\cdot p_2)(p_1\times p_2)\cdot\hat{z}\right]\;,
\end{eqnarray*}
$c=(p_1p_2)^{-1}$, and $\hat{z}$ is the unit vector perpendicular to
the ($xy$) plane. Here the $p_i$ ($i=x,y$) represent 2D wave vectors
that span the square $\{|p_x|,|p_y|<\Lambda\}$ in the Fourier space,
for which we have adopted the standard convention by defining
\begin{displaymath}
\int_p = \int_{-\Lambda}^\Lambda \int_{-\Lambda}^\Lambda d^2p
\end{displaymath}

Two coupling constants, $g_0=D_0/\lambda_0^2$ and
$g_\rho=D_\rho/\lambda_0^2$, appear in $S_e$, for which we carry out
an RG analysis in the limit, $\omega^2/\lambda_0\to 0$, in order to
derive, to one loop, the $\beta$ functions that describe their
behavior in the coupling space. Note that for those terms of $S_I$
with symmetric products of the fields under an exchange of momenta,
the corresponding coefficients will also retain the symmetric part.
For example, the coefficient of
$g_0\tilde{u}_1^R(p_1)u_1^R(p_2)\tilde{u}_1^R (p_3)u_1^R(p_4)$ is
written as a sum of the symmetric and antisymmetric parts,
\begin{widetext}
\begin{eqnarray}
& A(p_1,p_2)A(p_3,p_4)= &
\frac{1}{2}\left[A(p_1,p_2)A(p_3,p_4)+A(p_1,p_2)
A(p_3,p_2)\right]\;\nonumber\\
& & \ \ \ \ \
+\frac{1}{2}\left[A(p_1,p_2)A(p_3,p_4)-A(p_1,p_2)A(p_3,p_2)\right]\;,
\end{eqnarray}
\end{widetext} so that the antisymmetric part is cancelled by
integrating over the momenta.

\begin{center}
{\bf IV. RENORMALIZATION GROUP ANALYSIS}
\end{center}

To study whether the elastic waves are localized or delocalized in
the 2D heterogeneous media of the type that we consider, we apply
the RG method to the effective action, Eq. (\ref{action}). To do so,
we follow the momentum shell RG$^{29,30}$ and sum over the short
wavelength degrees of freedom. More specifically, we denote all the
fields in the action (\ref{action}) by $\Phi(k)$. To facilitate the
analysis, we change the domain of the integration from the square to
a circle of radius $\Lambda$. Since the small $k$ modes are supposed
to control the critical behavior of the system in the vicinity of
localization-delocalization transition, the change does not make any
qualitative difference to the results. Hereafter, we refer to the
small $k$ modes as the slow modes, and the rest as the fast modes.
We then define two sets of variables
\begin{eqnarray*}
\Phi_< &=& \Phi(k) \hspace{2mm} {\rm for} \hspace{2mm}
0<k<\Lambda/l\;,
\hspace{2mm} {\rm slow\; modes}\;,\\
\Phi_> &=& \Phi(k) \hspace{2mm} {\rm for} \hspace{2mm} \Lambda/l\leq
k \leq \Lambda\;, \hspace{2mm} {\rm fast\; modes}\;,
\end{eqnarray*}
where $l>1$ is the rescaling parameter of the RG transformation.
Then, the action is expressed in terms of $\Phi_<$ and $\Phi_>$ as
\begin{displaymath}
S(\Phi_<,\Phi_>)=S_0(\Phi_<)+S_0(\Phi_>)+S_I(\Phi_<,\Phi_>)\;.
\end{displaymath}

$S_0$ is a quadratic function of its arguments that can be separated
into slow and fast terms, but $S_I$ mixes the two modes. Then, the
partition function $Z$ is separated and written as follows
\begin{widetext}
\begin{displaymath}
Z=\int[{\cal D}\Phi_<] \int[{\cal D}\Phi_>]
\exp[S_0(\Phi_<)]\exp[S_0(\Phi_>)]\exp[S_I(\Phi_<,\Phi_>)]\equiv
\int[{\cal D}\Phi_<]\exp[S'_0(\Phi_<)]
\end{displaymath}
\end{widetext}
which defines the effective action $S'(\Phi_<)$ for the slow modes:
\begin{widetext}
\begin{eqnarray}\label{effective action}
\exp[S'(\Phi_<)] &=& \exp[S_0(\Phi_<)]\int[{\cal
D}\Phi_>]\exp[S_0(\Phi_>)]
\exp[S_I(\Phi_<,\Phi_>)]\nonumber\\
&=& \exp[S_0(\Phi_<)]\int[{\cal
D}\Phi_>]\exp[S_0(\Phi_>)]\frac{\int[{\cal D}
\Phi_>]\exp[S_0(\Phi_>)]\exp[S_I(\Phi_<,\Phi_>)]}
{\int[{\cal D}\Phi_>]\exp[S_0(\Phi_>)]}\nonumber\\
&=&
Z_{0>}\exp[S_0(\Phi_<)]\langle\exp[S_I(\Phi_<,\Phi_>)]\rangle_{0>}\;,
\end{eqnarray}
\end{widetext}
where $\langle\cdot\rangle_{0>}$ denotes an average with respect to
the fast modes, and $Z_{0>}$ is the partition function of
$S_0(\Phi_>)$ which adds a constant to the action, independent of
$\Phi_<$. The next step is to calculate the average
$\langle\exp[S_I(\Phi_<,\Phi_>)]\rangle_{0>}$, which we treat
perturbatively for weak disorder using the relation
\begin{eqnarray}\label{cumulant0}
\langle\exp(V)\rangle=\exp\left[\langle
V\rangle+\frac{1}{2!}(\langle V^2 \rangle-\langle
V\rangle^2)+\cdots\right]\;.
\end{eqnarray}
Therefore, according to Eqs. (\ref{effective action}) and
(\ref{cumulant0}), we have, up to one-loop order
\begin{eqnarray}\label{cumulant}
S'(\Phi_<)=\langle S_I\rangle+\frac{1}{2!}(\langle
S_I^2\rangle-\langle S_I \rangle^2)\;.
\end{eqnarray}
\begin{figure}[t]
\epsfxsize7truecm \epsffile{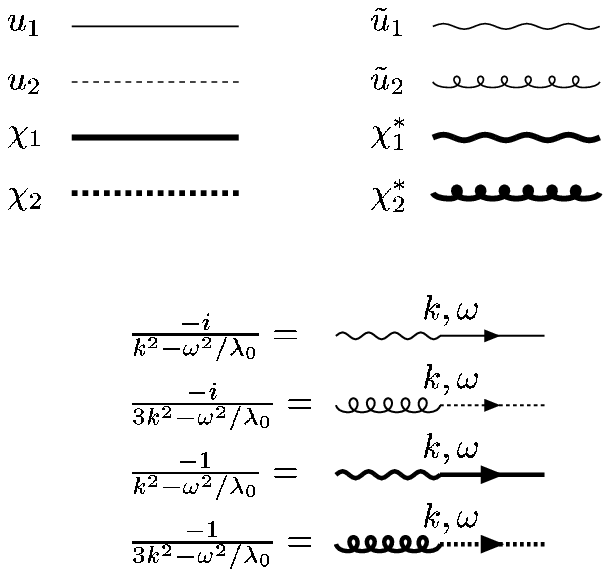}\caption{Graphical
representation of the fields and the Feynman rules for the
propagators.}
\end{figure}
\begin{figure}[]
\epsfxsize8truecm \epsffile{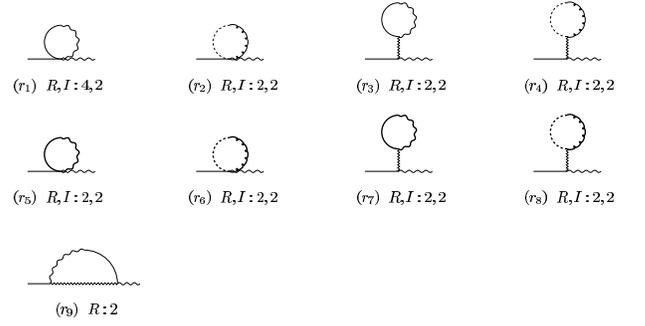}\caption{Feynman diagrams for
renormalization of the kinetic term $-i\tilde{u}^R_1(-k)k^2u^R_1(k)$
of $S_0$. They appear in the cumulant expansion to the lowest order.
External legs are the slow modes, while the internal fields are the
fast modes, and the integration is done over the fast modes. Those
fields that cosist of loops can be real or imaginary, and are
denoted by $R$ and $I$, respectively. The number of choices for the
construction of each diagram is also shown. The diagrams with
long-range interactions (zigzag lines) are divergent, due to the
zero momentum carried by the zigzag lines, but such diagrams are
canceled by the corresponding diagrams with Grassmanian loops (thick
lines). In fact, only $r_1$ and $r_9$ contribute.}
\end{figure}
Each term in the series contains some monomials in the fast and slow
modes. The former must be averaged with respect to $S_0(\Phi_>)$.
The first term in Eq. (\ref{cumulant}) yields {\it tree-level}
terms, as well as the corrections to the kinetic term of $S_0$. We
introduce a graphical representation of the terms which is shown in
the Fig. 1. The Feynman diagrams that contribute to the kinetic term
of the propagator $\tilde{u}_1^R u_1^R$ are shown in the Fig. 2.
According to Fig. 2, apart from a naive dimensional rescaling, one
should rescale the fields by a factor $F$ in the following way, in
order to keep the coefficient of the kinetic term to be the same as
in the original action
\begin{equation}
\Phi\rightarrow\frac{1}{\sqrt{F}}\Phi\;,
\end{equation}
with
\begin{eqnarray}\label{Z}
F = 1-18\pi(\Lambda^2g_0+\Lambda^{-2\rho+2}g_\rho)\delta l+{\cal
O}(g_0^2,g_\rho^2,g_0g_\rho)\;,\nonumber\\
\end{eqnarray}
where $\delta l=l-1$.

We now derive the RG equations for the disorder strengths by
renormalization of the coupling of the vertex
$\tilde{u}_1^R(p_1)u_1^R(p_2)\tilde{u}_1^R(p_3) u_1^R(p_4)$. The
Feynman diagrams that contribute to the renormalized coupling in
one-loop order of the perturbation expansion, and the corresponding
symmetry factors, are shown in the Fig. 3. Note that the couplings
are functions of the momenta and, therefore, we consider the first
term in the Taylor expansion and set the external momenta to be
parallel. The expressions for all the Feynman diagrams are listed in
the Appendix. It can be seen by dimensional analysis that the
canonical dimensions of the couplings in units of length are
\begin{eqnarray}
& & [g_0]=2\;, \\
& & [g_\rho]=2-2\rho.
\end{eqnarray}

\begin{figure}[t]
\epsfxsize8truecm \epsffile{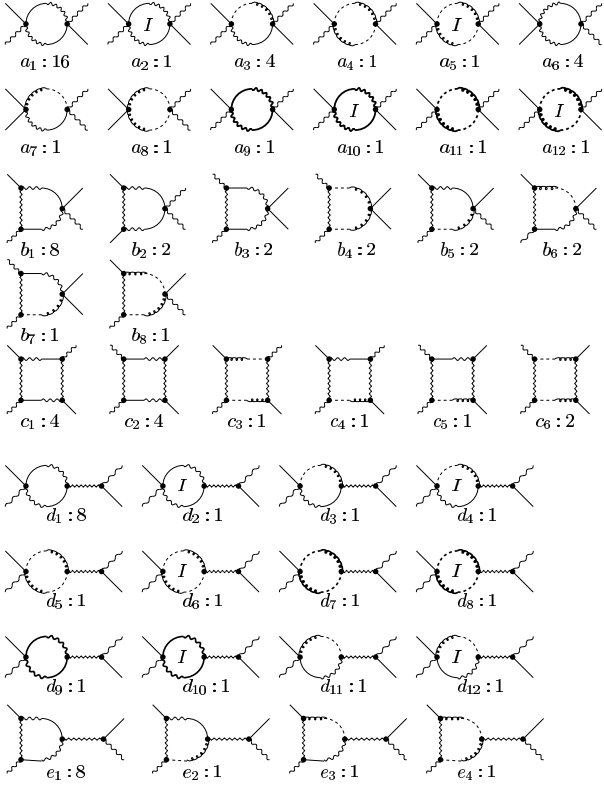}\caption{Feynman diagrams for
the renormalization of $\tilde{u}_1^R(p_1)
u_1^R(p_2)\tilde{u}_1^R(p_3)u_1^R(p_4)$ in the action. The diagrams
with loops of imaginary fields are indicated with $I$. The number of
choices for constructing each diagram is also given.}
\end{figure}

\begin{figure}[t]
\epsfxsize8truecm \epsffile{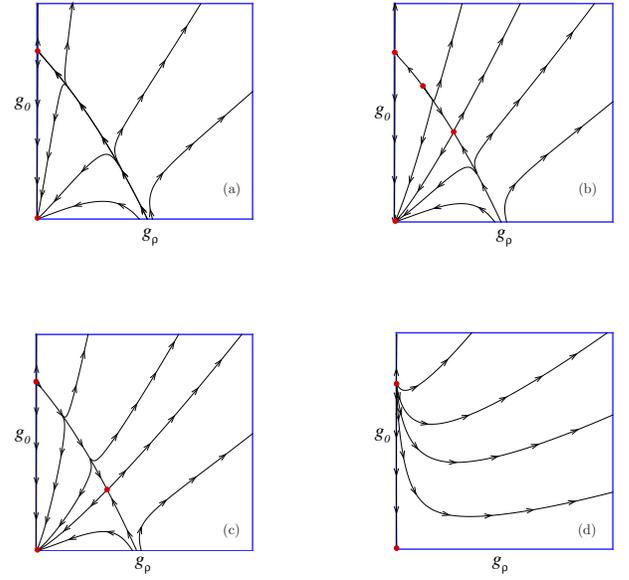}\caption{Renoemalization group
flows for (a) $\rho<0.14$; (b) $0.14<\rho<0.18$; (c) $0.18<\rho<1$,
and (d) $\rho>1$.}
\end{figure}

The following rules should be considered in expressing the Feynman
diagrams of the vertex function shown in Fig. 3:

(i) The diagrams that are made by different vertices have an extra
factor of 2, due to the quadratic term of the cumulant expansion.

(ii) All the diagrams have a factor $1/2!$ due to the cumulant
expansion.

(iii) An extra (-1) factor should be included for the diagrams with
Grassmanian loop.

\noindent Given the above rules, we obtain the following
renormalized couplings
\begin{eqnarray}
&&g'_0  l^2\Lambda'^2=F^{-2}\Big(g_0\Lambda^2+\frac{937}{18}\pi
g_0^2 \Lambda^4\delta l+\frac{716}{27}\pi
g_\rho^2\Lambda^{-4\rho+4}\delta l\nonumber\\&& \hspace{3cm} +
\frac{3635}{54}\pi g_0g_\rho\Lambda^{-2\rho+4}\delta l\Big )\;,\\
&&g'_{\rho}  l^{-2\rho+2}\Lambda'^{-2\rho+2}=F^{-2}\Big(g_\rho
\Lambda^{-2\rho+2}+36\pi g_0g_\rho\Lambda^{-2\rho+4}\delta
l\nonumber\\&&\hspace{4.5cm}+\frac{1948}{27}g_\rho^2\Lambda^{-4\rho+4}\delta
l\Big)\;,
\end{eqnarray}
where $\Lambda'=\Lambda/l$. Using Eq. (\ref{Z}) and
writing the equations in differential forms, we obtain the following
expressions for the $\beta$ functions that describe the couplings
\begin{eqnarray}\label{betag0}
\beta(\tilde{g}_0)\equiv\frac{\partial\tilde{g}_0}{\partial\ln
l}&=&-2\tilde{g}_0+\left(36+\frac{937}{18}\right)\tilde{g}_0^2+\frac{716}{27}\tilde{g}_\rho^2\nonumber\\
&&+
\left(36+\frac{3635}{54}\right)\tilde{g}_0\tilde{g}_\rho\;,\\
\beta(\tilde{g}_\rho)\equiv\frac{\partial\tilde{g}_\rho}{\partial\ln
l}&=&
(2\rho-2)\tilde{g}_\rho+72\tilde{g}_0\tilde{g}_\rho\nonumber\\&&+\left(36+\frac{1948}{27}
\right)\tilde{g}_\rho^2\;,
\end{eqnarray}


where $\tilde{g}_0$ and $\tilde{g}_\rho$ are dimensionless
parameters defined by
\begin{eqnarray}
& \tilde{g}_0 & =\pi g_0\Lambda^2\;,\\
& \tilde{g}_\rho & =\pi g_\rho\Lambda^{-2\rho+2}\;.
\end{eqnarray}

The $\beta$ functions that we have derived, Eqs. (\ref{betag0}) and
(\ref{betagrho}), describe how the two couplings - $g_0$ and
$g_\rho$ - behave, if we rescale all the lengths and consider the
elastic medium at coarser scales. If, for example, a small $g_0$
diverges under the RG rescaling, its implication is that a small
$g_0$ at small length scales behaves as very strong disorder at much
larger scales. Therefore, under such condition, every wave amplitude
will be localized. If, on the other hand, for some $g_0<g_c$ (where
$g_c$ is a critical value of $g_0$) $g_0$ vanishes under the RG
rescaling, it implies that, in this regime, $g_0$ does not
contribute much to the behavior of the propagating waves at large
length scales. Therefore, one way of defining a localized state may
be as follows: The waves are localized if, under the RG rescaling,
{\it at least} either $g_0$ or $g_\rho$ diverges.

We must also point out that, one may begin the RG rescaling and
analysis with the assumption that the couplings $g_0$ and $g_\rho$
are small. If, under the RG rescaling, we find stable fixed points
(FPs), it would imply that the assumption of the couplings being
small about such FPs is still valid. However, around an unstable FP
the couplings can grow and, hence, the perturbation expansion that
we have developed would fail. For our main purpose, however, namely,
determining the localized/extended regimes and the transition
between them, the most important goal is to determine the
condition(s) under which the FPs are unstable, around which the
couplings can diverge.

The FPs of the model are the roots of $\beta$ functions. The RG
equations, together with the parameter $\rho$, have a complex phase
space. Depending on $\rho$, there are four regimes:

(i) For $\rho<(-17557527+128\sqrt{19977620601})/3888601\simeq 0.14$
there are two FPs: The trivial Gaussian FP,
$\{\tilde{g}^*_0=\tilde{g}_\rho^*=0\}$, which is stable, and a
nontrivial FP, $\{\tilde{g}_0^*=36/1585\simeq 0.022,\;
\tilde{g}_\rho^*=0\}$ which has one positive eigenvalue (along the
eigendirection of which is unstable) and one negative one (along the
eigendirection direction of which is stable); see Fig. 4a.
Physically, this implies that the diagram is divided into two parts.
In one part the Gaussian FP is relevant and the disorder does not
have any effect, so that all the states are delocalized. In the
second part, the values of couplings increase under rescaling, so
that the disorder (both random and correlated) is relevant and,
therefore, the elastic waves are localized. Thus, the line (more
precisely, the curve) that separates the two parts is where the
localization-delocalization transition takes place.

(ii)  Four FPs exist if $0.14<\rho< 289/1585\simeq 0.18$. The
Gaussian FP is stable. The other FPs are unstable in one
eigendirection but stable in the other eigendirection, except,
$\{\tilde{g}_0^*=0.022,\;\tilde{g}_\rho^*=0\}$, which has positive
eigenvalues and, hence, is unstable in all directions. This is shown
in Fig. 4b.

(iii) There are three FPs for $0.18<\rho<1$. The Gaussian FP is
again stable. The FP, $\{g_0^*=0.022\;,g^*_\rho=0\}$ is unstable in
all directions The third FP is unstable in one eigendirection but
unstable in the second eigendirection. Figure 4c presents this part
of the RG flow diagram.

In both (ii) and (iii), as $\rho$ increases, the system tries to
move away from case (i) (the delocalized-localized transition) to a
purely localized state (see also below). Moreover, in (i) - (iii)
there is a point on the $g_\rho$ axis which obviously is not a FP,
but the RG flows change their direction on the $g_\rho$ axis at that
point. This means that one of the $\beta$ functions is zero on this
axis, while the other one is not.

(iv) For $\rho>1$ there are two FPs. As Fig. 4d indicates, the
Gaussian FP is stable on the $\tilde{g}_0$ axis but unstable on the
$\tilde{g}_\rho$ axis, and the nontrivial FP,
$\{\tilde{g}^*_0=0.022,\;\tilde{g}^*_\rho=0\}$, is unstable in all
directions. The implication is that, while the power-law correlated
disorder is relevant, no new FP exists to one-loop order and,
therefore, the long-wavelength behavior of the system is determined
by the long-range component of the disorder. This means that for
$\rho>1$ the elastic waves are localized in 2D.

Let us note that the extension of the present RG analysis to 3D
systems is difficult, but doable. The reason for the difficulty is
twofold. (i) It is difficult to determine the transformation matrix
{\bf U} [see Eq. (\ref{U})] for a 3D system, as its forms becomes
very complex in 3D. Knowledge of {\bf U} is necessary for
diagonalizing the relevant matrices. (iii) As the Appendix
indicates, the number of contribution Feynman diagrams is large is
in 2D. The number of such diagrams much larger in 3D.

\begin{center}
{\bf V. COMPARISON WITH ACOUSTIC WAVE PROPAGATION}
\end{center}

Since scalar equations have often been invoked for describing
propagation of elastic waves, it is of interest to compare the above
results with those that we derived previously$^{25,26}$ for the
scalar model of (acoustic) wave propagation in heterogeneous media
with precisely the same type of disorder as what we consider in the
present paper. The governing equation for such waves is given by
\begin{equation}\label{acoustic}
m\frac{\partial^2u}{\partial t^2}=\mbox{\boldmath$\nabla$}\cdot
[\lambda({\bf x})\mbox{\boldmath$\nabla$}u({\bf x})].
\end{equation}
The analysis was carried out$^{25,26}$ for a $d-$dimensional system,
but we summarize its results for 2D media. The RG analysis indicated
that, depending on $\rho$, there can be two distinct regimes (unlike
the four regimes described above):

(i) For $0<\rho<1$ there are three sets of FPs. One set represents
the Gaussian FP, $\{g^*_0=g_\rho^*=0\}$, which is stable. The other
two are $\{g_0^*=1/4,\;g_\rho^*=0\}$, and [31]
\begin{widetext}
\begin{eqnarray}
& & g_0^*=-\left[\frac{4}{11}d+\frac{5}{44}(2\rho-d)\right]
+\sqrt{\left[\frac{4}{11}d+\frac{5}{44}(2\rho-d)\right]^2+\frac{5}{167}
(2\rho-2)^2}\;, \nonumber\\
& & g_\rho^*=\frac{3}{4}g_0^*+\frac{1}{16}(d-2\rho)\;,
\end{eqnarray}
\end{widetext}
which, for $d=2$, reduces to
\begin{widetext}
\begin{eqnarray}
& &
g^*_0=-\frac{1}{22}(5\rho+11)+\sqrt{\left[\frac{8}{11}+\frac{5}{11}
(\rho-1)^2\right]^2+\frac{5}{167}(2\rho-2)^2}\;,\nonumber\\
& & g^*_\rho=\frac{3}{4}g_0^*+\frac{1}{8}(1-\rho)\;,
\end{eqnarray}
\end{widetext} which is stable in one eigendirection but unstable in
the other eigendirection. Therefore, for $0<\rho<1$ the one-loop RG
analysis indicated that a medium with uncorrelated disorder is
unstable against long-range correlated disorder towards a new FP in
the space of the coupling constants. Hence, there is a phase
transition from delocalized to localized acoustic waves with
increasing the disorder intensity.

Thus, the physical implication of the RG results for acoustic wave
propagation described by Eq. (\ref{acoustic}) is as follows. In the
interval, $0<\rho<1$, there is a region in the space of the coupling
constants $\{g_0,g_\rho\}$ in which the RG flows take any initial
point to the Gaussian FP. This implies that, for $0<\rho<1$, a
disordered medium of the type considered in this paper and our
previous work$^{25,26}$ looks like a pure (ordered or homogeneous)
medium at large length scales, implying that acoustic waves are {\it
extended} or delocalized.

However, when $g_0$ or $g_\rho$ are large enough that the initial
point is out of the basin of attraction of the Gaussian FP, the RG
flows move such points toward large values, hence implying that,
under the RG rescaling, the probability density function of the
disorder becomes broader and broader at increasingly larger length
scales. Therefore, in this case, a propagating acoustic wave samples
a medium with very large spatial fluctuations in the elastic
stiffness or moduli. We also found that,$^{25,26}$ even if one
starts in a disordered medium with purely long-range correlations
(i.e., one with $g_0=0$), the RG equations indicate that the growth
of $g_\rho$ will lead to increasing, i.e., {\it nonzero}, $g_0$,
hence implying that {\it uncorrelated} disorder will be produced by
the rescaling. Since the local fluctuations in the bulk moduli play
the role of scattering points, the implication for acoustic waves is
that the multiple scattering of a propagating wave from the
uncorrelated disorder will destroy the wave's coherence, leading
eventually to the localization of acoustic waves.

(ii) For $\rho>1$ there are two FPs: the Gaussian FP which is stable
on the $g_0$ axis but not on the $g_\rho$ axis, and a second FP,
$\{g^*_0=1/4,\; g^*_\rho=0\}$, which is unstable in all directions.
The implication for acoustic waves is that, although power-low
correlated disorder is relevant, no new FP exists to one-loop order
and, therefore, the system's long-wavelength behavior is determined
by the long-range component of the disorder. This implies that for
$\rho>1$ the acoustic waves are localized (in fact, in this case
they are localized for any $d$), which is similar to the elastic
waves studied in the present paper. In addition, in both cases the
system undergoes a disorder-induced transition when only the
uncorrelated disorder is present.

Let us note that we argued in our previous papers$^{25,26}$ that, in
the case of acoustic waves, although, similar to the elastic waves
considered in the present paper, the RG calculations were carried
out to one-loop order, the analysis should still be valid for higher
orders of the perturbation as well. The argument was based on the
fact that the signs of the higher-order terms are all positive. That
this is so is due to the following. We must keep in mind that the
contraction coefficients for auxiliary fields are always greater
than those of auxiliary {\it and} Grassmanian fields that supply the
negative terms. Moreover, the numbers of diagrams of, e.g., a real
auxiliary field and an imaginary auxiliary field are equal to number
of diagrams of an auxiliary {\it and} Grassmanian field. This
implies immediately that the signs of higher-order terms should also
be positive. We, therefore, concluded that$^{25,26}$ the one-loop
results for the acoustic waves should be valid to all orders.
However, we now believe that this is only a necessary but not
sufficient conditions. In the case of elastic waves, though, we
cannot even determine {\it a priori} the signs of the higher-order
terms.

Thus, comparison of propagation of elastic and acoustic waves in the
type of heterogeneous media that we consider in this paper indicates
that, while the RG flow diagrams for the elastic waves is more
complex than those of the acoustic waves, the region of the coupling
constants space in which they are delocalized is {\it narrower} than
that of the acoustic waves.

\begin{center}
{\bf VI. SUMMARY}
\end{center}

We developed a field-theoretic description of propagation and
localization of elastic waves in 2D heterogeneous solids using a RG
approach. Two types of heterogeneities, random disorder and one with
long-range correlations with a power-law correlation function, were
considered. We found that in presence of power-law correlated
disorder with the exponent $\rho>1$ (non-decaying correlations) the
RG flows are toward the strong coupling regime, and the waves are
localized. For $\rho<1$, and depending on its value, there are other
fixed points. One, which is stable, is the Gaussian FP with a small
domain of attraction. In this domain, long-range correlated
disorder, as well as the random disorder, are irrelevant and,
therefore, the waves are delocalized. In this regard, the
delocalized states in the Gaussian domain are unlike electrons in 2D
systems, which remain localized for any disorder.

Whether the delocalized states predicted for the Gaussian domain
persist, if we analyze the RG flows beyond the one-loop
approximation, remains to be seen. It may be that the domain of
attraction of the Gaussian FP shrinks (and might disappear
completely), if we consider the contributions of the higher order
loops. However, analytical determination of the contribution of even
the second-order loops for this problem is very difficult.

As we mentioned in the Introduction, a challenging feature of the
localization problem is obtaining an analytical estimate of the
localization length exponent. In this regard, the previous
analytical approaches are in contradiction with the numerical
results. We hope that the method developed in this paper can provide
a precise way of describing the critical properties of the
localization-delocalization transition and its critical exponents in
higher dimensions.

We are currently carrying out extensive numerical simulations in
order to further check the accuracy of the predictions of the
dynamical RG method developed in this paper. The results will be
reported in the near future.

\begin{center}
{\bf ACKNOWLEDGMENTS}
\end{center}

The work of R.S. was supported by the NIOC. We thank Mehdi Vaez
Allaei for his valuable help in the preparation of the manuscript.

\begin{center}
{\bf Appendix: Integrals for the Feynman Diagrams}
\end{center}

In this Appendix we list all the expressions for the Feynman
diagrams shown in Fig. 3.
\begin{widetext}
\begin{eqnarray*}
a_1 &=& \int \frac{d^2q}{(q^2-\frac{\omega^2}{\lambda_0})^2}
\times\frac{1}{4}[A(p_1,p_2)A(-q,q)+A(q,p_2)A(p_1,-q)]\\
& \times & [A(p_3,p_4)A(q,-q)+A(q,p_4)A(p_3,-q)]\;.\\
&=& \frac{1}{4}p_1p_2p_3p_4\int qdq \int d\theta
[9+(3\cos^2\theta+\sin^2\theta)^2]^2=\frac{1523\pi}{16}p_1p_2p_3p_4
\int qdq\;.\\
a_2 &=& \int\frac{d^2q}{(q^2-\frac{\omega^2}{\lambda_0})^2}\times
4[A(p_2,p_1)
A(q,-q)][A(p_4,p_3)A(q,-q)]\;.\\
&=& 324p_1p_2p_3p_4\int qdq \int d\theta=648\pi p_1p_2p_3p_4\int qdq\;.\\
a_3 &=&
-\int\frac{d^2q}{(3q^2-\frac{\omega^2}{\lambda_0})(q^2-\frac{\omega^2}
{\lambda_0})}\times\frac{1}{2}[A(-q,p_1)C(p_2,q)][A(p_4,q)C(-q,p_3)]\;.\\
&=& \frac{1}{3}p_1p_2p_3p_4\int qdq \int d\theta(3\cos^2\theta+
\sin^2\theta)^2\sin^2(2\theta)=\frac{17\pi}{12}p_1p_2p_3p_4\int qdq\;.\\
a_4 &=& \int\frac{d^2q}{(3q^2-\frac{\omega^2}{\lambda_0})^2}\times
4[A(p_2,p_1)B(-q,q)-C(p_2,q)C(-q,p_1)]\\
& \times & [A(p_4,p_3)B(-q,q)+C(p_4,q)C(-q,p_3)]\;.\\
&=& \frac{4}{9}p_1p_2p_3p_4\int qdq \int d\theta
[3+\sin^2(2\theta)]^2=11\pi p_1p_2p_3p_4\int qdq\;.\\
a_5 &=& \int\frac{d^2q}{(3q^2-\frac{\omega^2}{\lambda_0})^2}\times
4[A(p_2,p_1)
B(-q,q)][A(p_4,p_3)B(q,-q)]\;.\\
&=& 4p_1p_2p_3p_4\int qdq\int d\theta=8\pi p_1p_2p_3p_4\int qdq\;.\\
a_6 &=& \int\frac{d^2q}{(q^2-\frac{\omega^2}{\lambda_0})^2}
\times \frac{1}{4}[A(q,p_1)A(-q,p_3)+A(-q,p_1)A(q,p_3)]\\
& \times & [A(p_2,-q)A(p_4,q)+A(p_4,-q)A(p_2,q)]\;.\\
&=& p_1p_2p_3p_4\int qdq \int d\theta
(3\cos^2\theta+\sin^2\theta)^2=9\pi p_1p_2p_3p_4\int qdq\;.\\
a_7
&=&-\int\frac{d^2q}{(3q^2-\frac{\omega^2}{\lambda_0})(q^2-\frac{\omega^2}
{\lambda_0})}\times 4[A(q,p_1)C(-q,p_3)][A(p_2,-q)C(p_4,q)]\;.\\
&=& \frac{4}{3}p_1p_2p_3p_4\int qdq \int
d\theta=\frac{8\pi}{3}p_1p_2p_3
p_4\int qdq\;.\\
a_8 &=&
\int\frac{d^2q}{3(q^2-\frac{\omega^2}{\lambda_0})(3q^2-\frac{\omega^2}
{\lambda_0})}\times\frac{1}{4}[C(q,p_1)C(-q,p_3)+C(-q,p_1)C(q,p_3)]\\
& \times & [C(p_2,-q)C(p_4,q)+C(p_4,-q)C(p_2,q)]\;.\\
&=& \frac{1}{9}p_1p_2p_3p_4\int qdq \int d\theta
\sin^4(2\theta)=\frac{\pi}{12}p_1p_2p_3p_4\int qdq\;.\\
\end{eqnarray*}
\begin{eqnarray*}
a_9 &=& a_{10}=-\int\frac{d^2q}{(q^2-\frac{\omega^2}{\lambda_0})^2}
\times 4[A(p_2,p_1)A(q,-q)][A(p_4,p_3)A(q,-q)]\;.\\
&=& -324p_1p_2p_3p_4\int qdq\int d\theta
=-648\pi p_1p_2p_3p_4\int qdq\;.\\
a_{11} &=&
a_{12}=-\int\frac{d^2q}{(3q^2-\frac{\omega^2}{\lambda_0})^2}
\times 4[A(p_2,p_1)B(q,-q)][A(p_4,p_3)B(q,-q)]\;.\\
&=& -4p_1p_2p_3p_4\int qdq\int d\theta
=-8\pi p_1p_2p_3p_4\int qdq\;.\\
b_1 &=&
\int\frac{d^2q}{(q^2-\frac{\omega^2}{\lambda_0})^2}\times\frac{1}{2}
[A(p_2,-q)A(p_1,q)][A(p_4,p_3)A(q,-q)+A(q,p_3)A(p_4,-q)]\;.\\
&=& \frac{1}{2}p_1p_2p_3p_4\int q^{-2\rho+1}dq \int
d\theta[9+(3\cos^2\theta+
\sin^2\theta)^2](3\cos^2\theta+\sin^2\theta)^2\\
&=&\frac{551\pi}{8}p_1p_2p_3p_4\int q^{-2\rho+1}dq\;.\\ b_2 &=&
b_3=\int\frac{d^2q}{(q^2-\frac{\omega^2}{\lambda_0})^2}
\times\frac{1}{4}[A(p_2,q)A(p_4,-q)+A(p_4,q)A(p_2,-q)]\\
& \times & [A(p_1,q)A(p_3,-q)+A(p_1,-q)A(p_3,q)]\;.\\
&=& p_1p_2p_3p_4\int q^{-2\rho+1}dq \int
d\theta(3\cos^2\theta+\sin^2\theta)^2
=9\pi p_1p_2p_3p_4\int q^{-2\rho+1}dq\;.\\
b_4 &=&
\int\frac{d^2q}{3(q^2-\frac{\omega^2}{\lambda_0})(3q^2-\frac{\omega^2}
{\lambda_0})}\times\frac{1}{4}[C(q,p_1)C(-q,p_3)+C(-q,p_1)C(q,p_3)]\\
& \times & [C(p_2,-q)C(p_4,q)+C(p_4,-q)C(p_2,q)]\;.\\
&=& \frac{1}{9}p_1p_2p_3p_4\int q^{-2\rho+1}dq \int d\theta
\sin^4(2\theta)=\frac{\pi}{12}p_1p_2p_3p_4\int q^{-2\rho+1}dq\;.\\
b_5 &=&
-\int\frac{d^2q}{3(q^2-\frac{\omega^2}{\lambda_0})(q^2-\frac{\omega^2}
{\lambda_0})}\times[A(p_4,-q)C(q,p_3)][A(q,p_1)C(p_2,-q)]\;.\\
&=& \frac{2}{3}p_1p_2p_3p_4\int q^{-2\rho+1}dq \int d\theta
\sin^2(2\theta)(3\cos^2\theta+\sin^2\theta)^2=\frac{17\pi}{6}p_1p_2p_3p_4\int
q^{-2\rho+1}dq\;.\\
b_6 &=&
-\int\frac{d^2q}{(3q^2-\frac{\omega^2}{\lambda_0})(q^2-\frac{\omega^2}
{\lambda_0})}\times 2[A(q,p_3)C(p_4,-q)][A(p_2,-q)C(q,p_1)]\;.\\
&=& \frac{2}{3}p_1p_2p_3p_4\int q^{-2\rho+1}dq \int d\theta
\sin^2(2\theta)(3\cos^2\theta+\sin^2\theta)^2=\frac{17\pi}{6}p_1p_2p_3p_4\int
q^{-2\rho+1}dq\;.\\
b_7 &=&
-\int\frac{d^2q}{(3q^2-\frac{\omega^2}{\lambda_0})(q^2-\frac{\omega^2}
{\lambda_0})}\times 4[A(q,p_1)C(-q,p_3)][A(p_2,-q)C(p_4,q)]\;.\\
&=& \frac{4}{3}p_1p_2p_3p_4\int q^{-2\rho+1}dq \int d\theta
\sin^2(2\theta)(3\cos^2\theta+\sin^2\theta)^2=\frac{17\pi}{3}p_1p_2p_3p_4\int
q^{-2\rho+1}dq\;.\\
b_8 &=&
\int\frac{d^2q}{(3q^2-\frac{\omega^2}{\lambda_0})(3q^2-\frac{\omega^2}
{\lambda_0})}\times 4[A(p_4,p_3)B(q,-q)-C(q,p_3)C(p4,-q)][C(p_2,-q)
C(q,p_1)]\;.\\
&=& \frac{4}{9}p_1p_2p_3p_4\int q^{-2\rho+1}dq \int
d\theta[3+\sin^2(2\theta)]
\sin^2(2\theta)=\frac{15\pi}{3}p_1p_2p_3p_4\int q^{-2\rho+1}dq\;.\\
c_1 &=& \int\frac{d^2q}{(q^2-\frac{\omega^2}{\lambda_0})^2}
\times[A(q,p_1)A(p_2,-q)][A(p_3,-q)A(q,p_4)]\;.\\
&=& p_1p_2p_3p_4 \int q^{-4\rho+1}dq \int
d\theta(3\cos^2\theta+\sin^2\theta)^4
=\frac{227\pi}{4}p_1p_2p_3p_4\int q^{-4\rho+1}dq\;.\\
c_2 &=& \int\frac{d^2q}{(q^2-\frac{\omega^2}{\lambda_0})^2}
\times[A(q,p_1)A(-q,p_3)][A(p_2,-q)A(p_4,q)]\;.\\
&=& p_1p_2p_3p_4\int q^{-4\rho+1}dq \int
d\theta(3\cos^2\theta+\sin^2\theta)^4
=\frac{227\pi}{4}p_1p_2p_3p_4\int q^{-4\rho+1}dq\;.\\
c_3 &=& \int\frac{d^2q}{(3q^2-\frac{\omega^2}{\lambda_0})^2}
\times 4[C(q,p_1)C(p_2,-q)][C(p_4,-q)C(q,p_3)]\;.\\
&=& \frac{4}{9}p_1p_2p_3p_4\int q^{-4\rho+1}dq \int d\theta
\sin^4(2\theta)=\frac{\pi}{3}p_1p_2p_3p_4\int q^{-4\rho+1}dq\;.\\
\end{eqnarray*}
\begin{eqnarray*}
c_4 &=&
\int\frac{d^2q}{(3q^2-\frac{\omega^2}{\lambda_0})(q^2-\frac{\omega^2}
{\lambda0})}\times(-4)[A(q,p_1)C(p_2,-q)][A(p_4,-q)C(q,p_3)]\;.\\
&=& p_1p_2p_3p_4\times\frac{4}{3}\int q^{-4\rho+1}dq \int d\theta
(3\cos^2\theta+\sin^2\theta)^2\sin^2(2\theta)=\frac{17\pi}{3}p_1p_2p_3p_4\int
q^{-4\rho+1}dq\;.\\
c_5 &=&
\int\frac{d^2q}{(3q^2-\frac{\omega^2}{\lambda_0})(q^2-\frac{\omega^2}
{\lambda_0})}\times(-4)[A(q,p_1)C(-q,p_3)][A(p_2,-q)C(p_4,q)]\;.\\
&=& \frac{4}{3}p_1p_2p_3p_4\int q^{-4\rho+1}dq \int d\theta
(3\cos^2\theta+\sin^2\theta)^2\sin^2(2\theta)=\frac{17\pi}{3}p_1p_2p_3p_4\int
q^{-4\rho+1}dq\;.\\
c_6 &=&
\int\frac{d^2q}{(3q^2-\frac{\omega^2}{\lambda_0})(3q^2-\frac{\omega^2}
{\lambda_0})}\times[C(q,p_1)C(-q,p_3)][C(p_2,-q)C(p_4,q)]\;.\\
&=& \frac{1}{9}p_1p_2p_3p_4\int q^{-4\rho+1}dq \int d\theta
\sin^4(2\theta)=\frac{\pi}{12}p_1p_2p_3p_4\int q^{-4\rho+1}dq\;.\\
d_1 &=&
\int\frac{d^2q}{(q^2-\frac{\omega^2}{\lambda_0})^2}\times\frac{1}{2}
[A(p_2,p_1)A(-q,q)+A(-q,p_1)A(p_2,q)][A(p_4,p_3)A(q,-q)]\;.\\
&=& \frac{9}{2}p_1p_2p_3p_4\int qdq \int d\theta[9+(3\cos^2\theta
+\sin^2\theta)^2]=\frac{243\pi}{2}p_1p_2p_3p_4\int qdq\;.\\ d_2 &=&
\int\frac{d^2q}{(q^2-\frac{\omega^2}{\lambda_0})^2}
\times 4[A(p_2,p_1)A(-q,q)][A(p_4,p_3)A(q,-q)]\;.\\
&=& 324p_1p_2p_3p_4\int qdq \int d\theta=324\pi p_1p_2p_3p_4\int qdq\;.\\
d_3 &=& d_4=0\;.\\
d_5 &=& d_6=\int\frac{d^2q}{(3q^2-\frac{\omega^2}{\lambda_0})^2}
\times 4[A(p_2,p_1)B(q,-q)][A(p_3,p_4)A(q,-q)]\;.\\
&=& 4p_1p_2p_3p_4\int qdq \int d\theta=8\pi p_1p_2p_3p_4\int qdq\;.\\
d_7 &=&
d_8=-\int\frac{d^2q}{(3q^2-\frac{\omega^2}{\lambda_0})^2}\times 4
[A(p_2,p_1)B(q,-q)][A(p_4,p_3)A(-q,q)]\;.\\
&=& -324p_1p_2p_3p_4\int qdq \int d\theta=-648\pi p_1p_2p_3p_4\int qdq\;.\\
d_9 &=& d_{10}=0\;.\\
e_1 &=& \int\frac{d^2q}{(q^2-\frac{\omega^2}{\lambda_0})^2}
\times[A(q,p_1)A(p_2,-q)][A(p_4,p_3)A(q,-q)]\;.\\
&=& 9p_1p_2p_3p_4\int q^{-2\rho+1}dq \int d\theta(3\cos^2\theta+
\sin^2\theta)^2=81\pi p_1p_2p_3p_4\int q^{-2\rho+1}dq\;.\\
e_2 &=& e_3=0\;.\\
e_4 &=& \int\frac{d^2q}{(3q^2-\frac{\omega^2}{\lambda_0})^2}
\times(-4)[C(q,p_1)C(p_2,-q)][A(p_4,p_3)B(q,-q)]\;.\\
&=& \frac{4}{3}p_1p_2p_3p_4\int q^{-2\rho+1}dq \int d\theta
\sin^2(2\theta)=\frac{4\pi}{3}p_1p_2p_3p_4\int q^{-2\rho+1}dq\;.
\end{eqnarray*}
\end{widetext}

\begin{center}
\line(1,0){230}
\end{center}

\noindent\textsuperscript{*}moe@iran.usc.edu

\begin{description}

\item $^1$P. W. Anderson, Phys. Rev. {\bf 109}, 1492 (1958); N. F. Mott and W.
D. Twose, Adv. Phys. {\bf 10}, 107 (1961).

\item $^2$S. He and J. D. Maynard, Phys. Rev. Lett. {\bf 57}, 3171 (1986); J.
D. Maynard, Rev. Mod. Phys. {\bf 73}, 401 (2001).

\item $^3$D. S. Wiersma, P. Bartolini, A. Lagendijk, and R. Righini, Nature
(London) {\bf 390}, 671 (1997).

\item $^4$E. Abrahams, P. W. Anderson, D. C. Licciardello, and T. V.
Ramakrishnan, Phys. Rev. Lett. {\bf 42}, 673 (1979); P. W. Anderson,
E. Abrahams, and T. V. Ramakrishnan, {\it ibid.} {\bf 43}, 718
(1979).

\item $^5$D. Vollhardt and P. W\"olfle, Phys. Rev. B {\bf 22}, 4666 (1980).

\item $^6$T. R. Kirkpatrick, Phys. Rev. B {\bf 31}, 5746 (1985); C. A. Condat
and T. R. Kirkpatrick, {\it ibid.} {\bf 33}, 3102 (1986); {\it
ibid}. {\bf 36}, 6782 (1987).

\item $^7$F. J. Wegner, Z. Phys. B {\bf 25}, 327 (1976); {\bf 35}, 207 (1979);
{\bf 36}, 209 (1980); Nucl. Phys. B {\bf 180}[FS2], 77 (1981).

\item $^8$K. B. Efetov, A. I. Larkin, and D. E. Khmelnitskii, Sov. Phys. JETP
{\bf 52}, 568 (1980).

\item $^9$S. Hikami, Phys. Rev. B {\bf 24}, 2671 (1981).

\item $^{10}$A. MacKinnon and B. Kramer, Z. Phys. B {\bf 53}, 1 (1983); B.
Kramer and A. MacKinnon, Rep. Prog. Phys. {\bf 56}, 1469 (1993).

\item $^{11}$P. C. Martin, E. D. Siggia, and H. A. Rose, Phys. Rev. A {\bf 8},
423 (1973).

\item $^{12}$N. Bleistein, J. K. Cohen, and J. W. Stockwell, Jr., {\it
Mathematics of Multidimensional Seismic Imaging, Migration, and
Inversion} (Springer, New York, 2001); A. Ishimaru, {\it Wave
Propagation and Scattering in Random Media} (Oxford University
Press, Oxford, 1997).

\item $^{13}$For recent experimental observation of weak localization of
seismic waves see, for example, E. Larose, L. Margerin, B. A. van
Tiggelen, and M. Campillo, Phys. Rev. Lett. {\it 93}, 048501 (2004).

\item $^{14}$S. John, H. Sompolinsky, and M. J. Stephen, Phys. Rev. B {\bf 27},
5592 (1983).

\item $^{15}$M. Sahimi, {\it Heterogeneous Materials I} (Springer, New York,
2003), chapters 6 and 9; P. Sheng, {\it Introduction to Wave
Scattering, Localization, and Mesoscopic Phenomena} (Academic, San
Diego, 1995).

\item $^{16}$For reviews see, for example, Refs. [15] and, T. Nakayama, K.
Yakubo, and R. Orbach, Rev. Mod. Phys. {\bf 66}, 381 (1994).

\item $^{17}$M. Belhadi, O. Rafil, R. Tigrine, A. Khater, A. Virlouvet, and K.
Maschke, Eur. Phys. J. B {\bf 15}, 435 (2004); J. J. Ludlam, S. N.
Taraskin, and S. R. Elliot, Phys. Rev. B {\bf 67}, 132203 (2003); J.
J. Ludlam, T. O. Stadelmann, S. N. Taraskin, and S. R. Elliot, J.
Non-Cryst. Solids {\bf 293-295}, 676 (2001); D. Garc\'ia-Pablos, M.
Sigalas, F. R. Montero de Espinosa, M. Torres, M. Kafesaki, and N.
Garcia, Phys. Rev. Lett. {\bf 84}, 4349 (2000).

\item $^{18}$B. Garber, M. Cahay, and G. E. W. Bauer, Phys. Rev. B {\bf 62},
12831 (2000).

\item $^{19}$F. M. Li, Y. S. Wang, C. Hu, and W. H. Huang, Waves in Random
Media {\bf 14}, 217 (2004).

\item $^{20}$B. A. van Tiggelen, L. Margerin, and M. Campillo, J. Acous. Soc.
Am. {\bf 110}, 1291 (2001).

\item $^{21}$L. Margerin, M. Campillo, and B. van Tiggelen, J. Geophys. Res.
{\bf 105}, 7873 (2000).

\item $^{22}$See, for example, M. Bouchon, Pure Appl. Geophys. {\bf 160}, 445
(2004); W. Sun and H. Yang, Acta Mech. Solida Sinica {\bf 16}, 283
(2004); T.-K. Hong and B. L. N. Kennet, Geophys. J. International
{\bf 154}, 483 (2003); {\it ibid.} {\bf 150}, 610 (2002); T. Bohlen,
Comput. Geosci. {\bf 28}, 887 (2002); H. A. Friis, T. A. Johansen,
M. Haveraaen, H. Muthe-Kaas, and A. Drottning, Appl. Num. Math. {\bf
39}, 151 (2001).

\item $^{23}$M. Sahimi and S. E. Tajer, Phys. Rev. E {\bf 71}, 046301 (2005).

\item $^{24}$M. R. Rahimi Tabar, M. Sahimi, F. Ghasemi, K. Kaviani, M.
Allamehzadeh, J. Peinke, M. Mokhtari, M. Vesaghi, M. D. Niry, A.
Bahraminasab, S. Tabatabai, S. Fayyazbakhsh, and M. Akbari, in {\it
Modelling Critical and Catastrophic Phenomena in Geoscience}, edited
by P. Bhattacharyya and B. K. Chakrabarti (Springer, Berlin, 2006).

\item $^{25}$F. Shahbazi, A. Bahraminasab, S. M. Vaez Allaei, M. Sahimi, and
M. R. Rahimi Tabar, Phys. Rev. Lett. {\bf 94}, 165505 (2005).

\item $^{26}$A. Bahraminasab, S. M. Vaez Allaei, F. Shahbazi, M. Sahimi, M. D.
Niry, and M. R. Rahimi Tabar, Phys. Rev. B {\bf 75}, 064301 (2007).

\item $^{27}$C. De Dominicis and L. Peliti, Phys. Rev. Lett. {\bf 38}, 505
(1977); Phys. Rev. B {\bf 18}, 353 (1978).

\item $^{28}$D. Hochberg, C. Molina, J. Perez, and M. Visser, Phys. Rev. E
{\bf 60}, 6343 (1999).

\item $^{29}$J. Cardy, {\it Scaling and Renormalization in Statistical Physics}
(Cambridge University Press, London, 1996).

\item $^{30}$N. Goldenfeld, {\it Lectures on Phase Transitions and the
Renormalization Group}\\ (Addison-Wesley, New York, 1992).

\item $^{31}$The formula that was given by Bahraminasab {\it et al.}$^{26}$
for $g_0^*$ contains some minor errors in the numerical
coefficients. Equation (32) provides the correct formula.

\end{description}

\end{document}